\documentclass[authoryear,12pt]{elsarticle}

\newenvironment{acks}[1][Acknowledgements]{\footnotesize\paragraph*{#1}}{}
\def\arcsec{$^{\prime\prime}$}
\pdfoutput=1

\usepackage{amssymb}

\usepackage[
a4paper=true,
breaklinks=true,
colorlinks=true,
pdfauthor={D.Y.Kolobov, N.I. Kobanov, A.A. Chelpanov, A.A. Kochanov, S.A. Anfinogentov, S.A. Chupin, I.I. Myshyakov, V.E. Tomin},
pdftitle={Behaviour of oscillations in loop structures above active regions}
]{hyperref}
\usepackage{textcomp}
\usepackage{gensymb}
\journal{Advances in Space Research}

\begin{document}

\begin{frontmatter}



\title{Behaviour of oscillations in loop structures above active regions}

\author[label1,label2]{D.Y.\,Kolobov\corref{cor}}
\ead{kolobov@iszf.irk.ru}
\cortext[cor]{Corresponding author}
\author[label1,label2]{N.I.\,Kobanov}
\author[label1,label2]{A.A.\,Chelpanov}
\author[label1,label2]{A.A.\,Kochanov}
\author[label1,label2]{S.A.\,Anfinogentov}
\author[label1,label2]{S.A.\,Chupin}
\author[label1,label2]{I.I.\,Myshyakov}
\author[label1,label2]{V.E.\,Tomin}

\address[label1]{Russia, 664033, Irkutsk, Lermontov St., 126-a}
\fntext[label2]{Institute of Solar-Terrestrial Physics, SB RAS}

\begin{abstract}
In this study we combine the multiwavelength ultraviolet -- optical
(Solar Dynamics Observatory, SDO) 
and radio (Nobeyama Radioheliograph, NoRH) observations to
get further insight into space-frequency distribution of
oscillations at different atmospheric levels of the Sun. We processed the 
observational data on NOAA~11711 active region and found oscillations 
propagating from the photospheric level through the transition region upward into the 
corona. The power maps of low-frequency (1--2 mHz) oscillations reproduce well
the fan-like coronal structures visible in the Fe \textsc{ix} 171\,\AA\ line.
High frequency oscillations (5--7 mHz) propagate along the vertical magnetic
field lines and concentrate inside small-scale elements in the umbra and at the 
umbra-penumbra boundary. We investigated the dependence of the dominant 
oscillation frequency upon the distance from the sunspot barycentre to
estimate inclination of magnetic tubes in higher levels of sunspots
where it cannot be measured directly, and found that this angle is close 
to 40\degree\ above the umbra boundaries in the transition region.
\end{abstract}


\end{frontmatter}

\section{Introduction}
The waves in solar magnetic structures play an important role in
processes of the energy exchange.
Moreover, the properties of the medium, i.e. the solar atmosphere, can be determined
by studying the wave propagation. This affects such problems as the origin of the solar 
magnetic fields, coronal heating, flares, wave propagation in plasma with different magnetic
field topology, and others. From the observation driven point of view the problems above 
depend on our knowledge of the magnetic field distribution with height, plasma velocity,
temperature and their variations. Because direct measurement of these parameters is 
a considerable challenge, in practice one has to use different tools to get a comprehensive 
picture of the atmosphere structure in active regions and processes therein.
The seismological study of the active regions is one of such unique tools
\citep{bogdan2006a,dem2012RSPTA}.
Different methods were developed to explore magnetic geometry of the sunspots and
to estimate the magnetic field inclination angle in the transition region and in the corona 
\citep{jess2013,yuan2014aa,kob2015sp}. Coronal seismology provides a capability to determine
the magnetic field strength in coronal closed magnetic structures \citep{nak2001aa}. 
Another opportunity is radio magnetography \citep{gel2004ASSL}. 
The radio emission is naturally connected with magnetic field providing an instrument to study the
3D-structure of the sunspot atmosphere. Joint observations in radio and optical spectral ranges 
have a great potential \citep{kis2006sp}, that has not been implemented yet. 
One can trace the wave train propagation and utilise magnetic field extrapolation to the corona 
to estimate the effective height of the radio emission, which can be used to qualify
the formation heights of the lines in UV and optical spectral ranges.
New facilities like the Siberian Solar Radio Telescope (SSRT,
\citealt{les2012sp}), the Chinese Spectral Radioheliograph (CSRH, \citealt{yan2013IAUS}), and others will provide new possibilities for observing the Sun's 
radio emission in the multifrequency mode.
\par Sunspot oscillations were extensively studied in the past decades (see reviews by 
\cite{lites1992a,sol2003,bogdan2006a,jess2015a}). Recent studies are based on multiwavelength 
observations and focus on determining the wave propagation speed
\citep{fel2010,abramov2011,kid2012sp,cha2015mnras}, wave dynamics \citep{sych2012,sych2014aa,zhu2014}, 
and related phenomena like jets, penumbral filaments, umbral flashes \citep{yur2014,yur2015,poz2015,
mad2015,luis2014,mau2013sp}, light bridges \citep{yuan2014apj,yang2015a} and also flares \citep{sych2014arXiv}.
Due to the circular symmetry, sunspots provide an opportunity to
study the wave propagation processes in the presence of magnetic fields. In sunspot umbra,
five-minute oscillations dominate at the photospheric level, and three-minute
ones prevail in the chromosphere \citep{kob1990sp,kentischer1995,sigwarth1997,rouppe2003,kobkol2009,RezSh2012}. 
Oscillations with longer periods (10--15 minutes) manifest themselves at the peripheral part 
of the sunspots. Other magnetic field regions, e.g. facula show very different behaviour
of the spatial-frequency dependence \citep{kobpul2007,cen2009}. Detecting and studing the coronal 
counterparts of these waves and quasi-periodic pulsations are relevant for the problem of energy 
flux transfer to the upper atmosphere.
\par We continue studing the space-frequency stratification of the 
oscillation processes in the sunspot atmosphere \citep{kob2015sp}. We extend the analysis 
with new sunspot data, including those obtained in radio emission, and compare the results
with magnetic field extrapolation.

\section{Observational Data and Methods}
\begin{figure}
\centerline{
\includegraphics[]{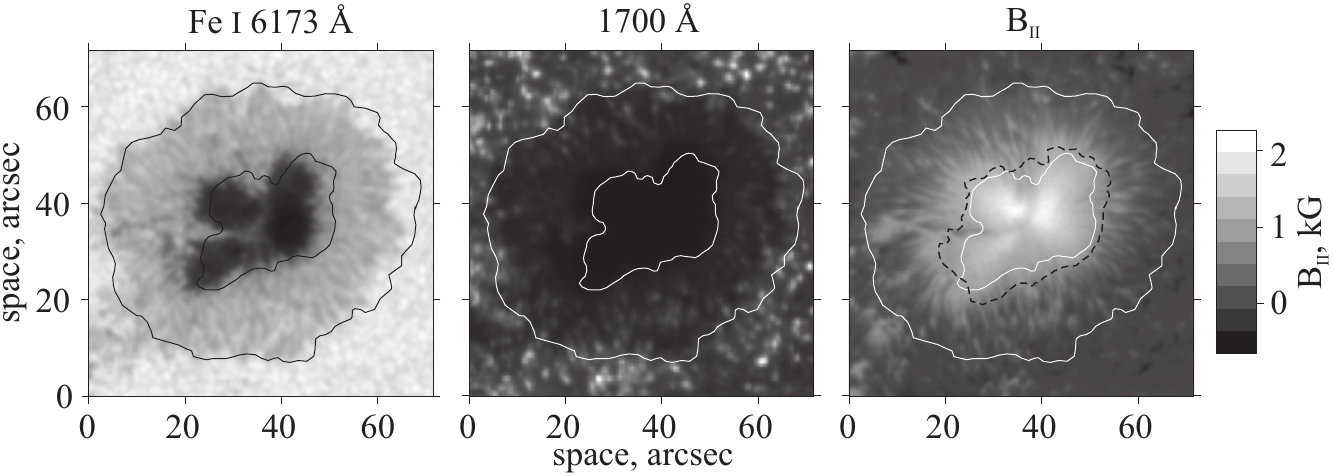}
}
\caption{Sunspot NOAA11711. Left panel: SDO/HMI continuum image.
Middle panel: the sunspot in the SDO/AIA 1700\,\AA\ band pass. Right
panel: magnetic field normal projection derived from the SDO/HMI vector
data. The inner and outer penumbra boundaries are outlined by black and white solid lines.
The dashed isoline marks the regions where the magnetic field
inclines at 45° to the surface normal.
}
\label{fig:brief}
\end{figure}
The paper is result of the investigation into the sunspot oscillations observed at
different wavelengths representing several heights of the solar atmosphere
from the photosphere to the corona. We used the data from two observatories 
in this study: Solar Dynamics Observatory (SDO, \cite{lem2012sp,sch2012sp,woods2012sp}) 
and Nobeyama Radioheliograph (NoRH, \citealt{Nakajima1994}).
The data were selected to cover height scale from the photosphere to the corona.

The sunspot had regular circular shape (\autoref{fig:brief}, left panel).
The observation was performed on 2013 April 6 00:00--04:20\,UT and
there was no flare activity at this period; a C1.7 flare occurred
on 10:58\,UT.

\par We chose four spectral bands provided by the Atmospheric Imaging Assembly (AIA)
for the analysis: Fe\,\textsc{ix} 171\,\AA\ and Fe\,\textsc{xii},
\textsc{xxiv} 193\,\AA\ (coronal heights), He\,\textsc{ii} 304\,\AA\
(transition region) and 1700\,\AA\
(upper photosphere, \autoref{fig:brief}, middle panel).
The Helioseismic and Magnetic Imager (HMI) data were
used to obtain the magnetic field at the photospheric level in
the Fe\,\textsc{i} 6173\,\AA\ line, which is formed at the 100\,--\,150\,km height
\citep{fleck2011sp,parnell1969sp}.
The vector magnetic-field components were computed using the code of
Very Fast Inversion of the Stokes Vector \citep{borrero2011sp}.
The SDO data sets were prepared using the \texttt{aia\char`_prep} and
\texttt{hmi\char`_prep} SolarSoft routines.
The sunspot was observed near the central meridian (W03S17), which allowed us to
minimize the projection effect on the analysis of oscillations at different heights.
We corrected the magnetic field angle to the solar normal within the
planar approximation using equation \citep{GaryHag1990}:
\begin{equation}
\varphi={cos}\gamma{cos}\varphi^\prime+{sin}\gamma{sin}\varphi^\prime{cos}\theta
\label{eq:magincl}
\end{equation}
where $\gamma$ is the angle between the line of sight and the solar normal; 
$\varphi^\prime$ is the observed field inclination angle; $\theta$ is the azimuthal angle corrected
for the line-of-sight deviation from the central-meridian plane
(\autoref{fig:brief}, right panel).

\par The radio data are the sequence of the 17 GHz NoRH images with the 20-second 
effective cadence and the 5\arcsec\ spatial resolution. The images were synthesised
with the ``Koshix'' algorithm on the Solar Data Analysis System. To increase the 
signal-to-noise ratio, the number of integration frames for synthesis was set to 10. The images 
were cropped to include the sunspot area ($400\times400$\arcsec). Then, the images were
co-aligned using the cross-correlation technique and smoothed over spatial dimensions.
The maximum brightness temperature in the sunspot does not exceed 29000\,K
and the degree of polarisation is about 25\%. Thus, the polarised
emission at 17\,GHz in this image is most likely due to optically thin
thermal bremsstrahlung. The line-of-sight magnetic field was estimated 
by the method proposed by \cite{gel2004ASSL}.

\par To study space-height properties of the sunspot oscillations we used
the method described in \cite{kob2015sp} that is suitable for circular
sunspots. To calculate the cut-off frequency, we used the equation 
\citep{bel1977,mcinjef2006,botha2011}:
\begin{equation}
f_c=\frac{g_0\gamma \mbox{cos}\varphi(r)}{4\pi v_s}
\label{eq:fc}
\end{equation}
where $f_c$ is the cut-off frequency; $g_0$=274 $\mbox{m\,s}^{-2}$ is the gravitational constant;
$v_s$ is the speed of sound; $\gamma=\frac{5}{3}$ is the
adiabatic index; $\varphi(r)$ is the dependence of inclination angle on
the barycenter distance.
Using \autoref{eq:fc} and empirical ratio $f_c=$0.81F(r)\,mHz$-1$\,mHz
for the 304\,\AA\ line \citep{yuan2014aa} we estimate 
the magnetic-field inclination angle $[\varphi]$ (\cite{kob2015sp}):
\begin{equation}
\varphi = \mbox{arccos}\left(\frac{4\pi v_s F(r)}{g_0\gamma} \right)
\label{eq:fi}
\end{equation}

\par The oscillations' power spectrum was calculated using the Fast Fourier Transform (FFT).
To remove trends from the data we subtracted the time series smoothed over a 
25-minute time interval from the original series.
The data set temporal resolution
varies from 12 to 45\,s for different instruments and bandpasses, which allows
us to analyse frequencies up to 11\,mHz --- the Nyquist frequency for the HMI
data; the Nyquist frequencies for the AIA data are 21 mHz (the 1700\,\AA\ band) and 
42\,mHz (the 304\,\AA\ and 171\,\AA\ bands), and that for the NoRH data is 25\,mHz. 
To analyse the spatial distribution of the dominant frequencies, we obtain FFT power maps 
in the following way. FFT power spectra were calculated
for every pixel ($0.6\times0.6$ arcsec$^2$) of the SDO/AIA/HMI intensity images. 
The calculated Fourier spectra were smoothed by convolution with a rectangular window 
1~mHz wide. The frequency corresponding to the power spectrum maximum value 
was taken as the dominant frequency for the particular spatial domain.
To analyse wave trains and to measure their propagation speeds the frequency filtration was 
performed by the sixth-order Morlet wavelet.
We used the algorithm described by \cite{torrence1998} to derive the FFT power spectra,
to wavelet-filter the data and estimate the statistical significance.
The normalisation by $1/\sigma^2$ gives a measure of the power relative to the white noise
level: 3$\sigma^2$ correspond to the 95\% confidence level, and 4.6$\sigma^2$ --- 99\%
confidence level ($\sigma^2$ is the variance of a time series).

\par The coronal magnetic field was calculated in the potential and nonlinear 
force-free (NLFF) approach, using SDO/HMI vector magnetograms as input data. To remove
180\degree\ ambiguity in the direction of filed component transversal to the line-of-sight 
we used the SFQ method by \cite{rud2014sp}. The potential field was calculated using the method,
described in \cite{alis1981aa} by applying FFT. The NLFF reconstruction was performed 
by using the optimization method, proposed by \cite{whe2000apj}. This method implementation 
by \cite{rud2009sp} was used in the present study: we take into account a full set of 
the evolutionary equations and allow the field variations not only in the inner volume but
also on the lateral and top boundaries of the computational domain.

\begin{figure}
\centerline{
\includegraphics[]{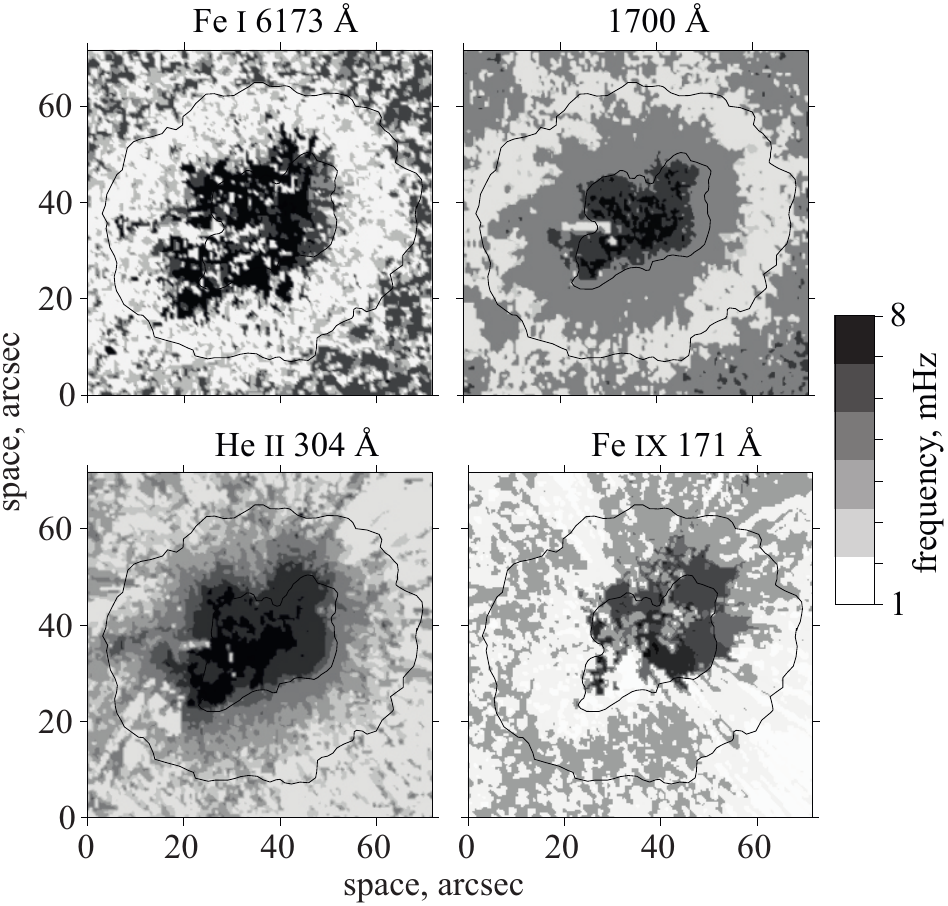}
}
\caption{Spatial distributions of dominant frequencies from the photosphere to
the corona. The black solid lines mark the inner and outer penumbral
boundaries.
        }
\label{fig:frequency_distributions}
\end{figure}

\begin{figure}
\centerline{
\includegraphics[]{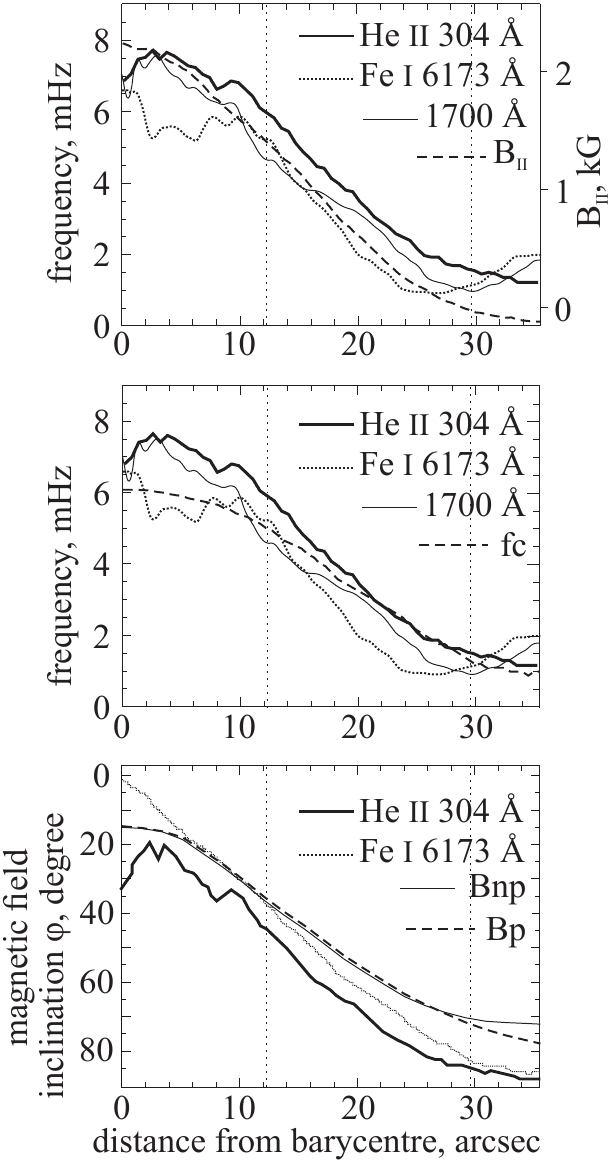}
}
\caption{Upper panel: dominant frequencies at three heights, and
the magnetic field as a function of distance from the sunspot
barycentre. Middle panel: dominant frequencies and the cut-off
frequency derived from \autoref{eq:fc}.
Bottom panel: magnetic field inclination angle derived from
\autoref{eq:fi} for the He\,\textsc{ii} 304\,\AA\
level, HMI Fe\,\textsc{i} 6173\,\AA\ data,
potential ($B_p$) and nonlinear force-free ($B_{np}$) magnetic field extrapolation 
at the 2200\,km level.
        }
\label{fig:barycentre}
\end{figure}

\section{Results}
\autoref{fig:frequency_distributions} shows the spatial distribution of 
dominant frequencies in the sunspot at different heights from the photosphere 
to the corona. The black solid lines mark the inner and outer penumbra boundaries.
Five-minute oscillations concentrate in an annular area,
that expands beyond the umbra boundaries with increasing height. Low-frequency
oscillations appear in the lower photosphere of the penumbra. Such a topology
resembles the concept explaining running penumbral waves as upwardly
propagating waves that form a “visual pattern”
\citep{rouppe2003,bloomfield2007a,kobkol2009}.
Low-frequency oscillations occupy the larger area in the picture plane
at higher coronal level visible in Fe\,\textsc{ix} 171\,\AA:
the areas of dominant frequencies become elongated in the radial direction.
Using the circular-shaped power distribution, one can
plot frequency vs. distance from the barycentre for different
height levels observed with AIA and HMI. \autoref{fig:barycentre} 
presents a dominant frequency [$F(r)$]
averaged over the pixels located at distance $r$ from the sunspot barycentre with a
0.6\arcsec\ step. 
The same averaging
was used to obtain
the longitudinal magnetic field [$B_{\shortparallel}$] and the absolute
value of the magnetic-field inclination (\autoref{fig:barycentre}).
We estimate the cut-off frequency $f_c$ (\autoref{fig:barycentre}, middle panel) and
the magnetic-field inclination angle $[\varphi]$ (\autoref{fig:barycentre}, bottom panel)
by Equations \ref{eq:fc} and \ref{eq:fi}.
Using potential $B_p$ and nonlinear force-free $B_{np}$ magnetic field 
extrapolation data we produce $[\varphi]$ vs. distance from barycentre 
plots for the 2200\,km level. Both cases show a very similar dependence 
except for the outer penumbra 
(\autoref{fig:barycentre}, bottom panel).
In the \autoref{fig:barycentre} middle panel one can see a  penumbra region, 
in which the dominant frequency curve (304\,\AA) follows the cut-off 
frequency curve. Earlier, \cite{kur2009} showed that pressure impulses engendered 
in the photosphere can excite oscillations at the cut-off frequency. Consequently, 
the oscillations that coincide with the resonant frequencies of the cavities under
the magnetic canopy amplify and can penetrate into the upper layers of the solar
atmosphere \citep{sri2008}.

\begin{figure}
\centerline{
\includegraphics[]{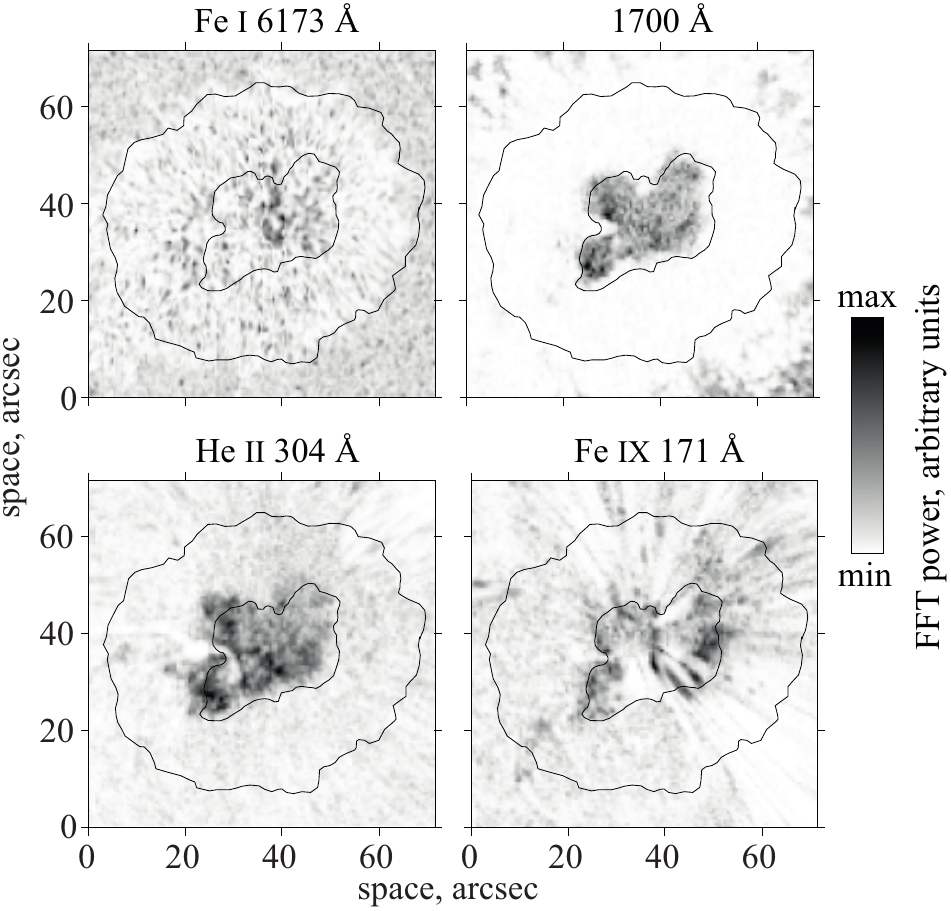}
}
\caption{Intensity FFT power maps averaged in the 6--7 mHz band. 
The black solid lines mark the inner and outer penumbral boundaries.
} 
\label{fig:ht-6-7}
\end{figure}

\begin{figure}
\centerline{
\includegraphics[]{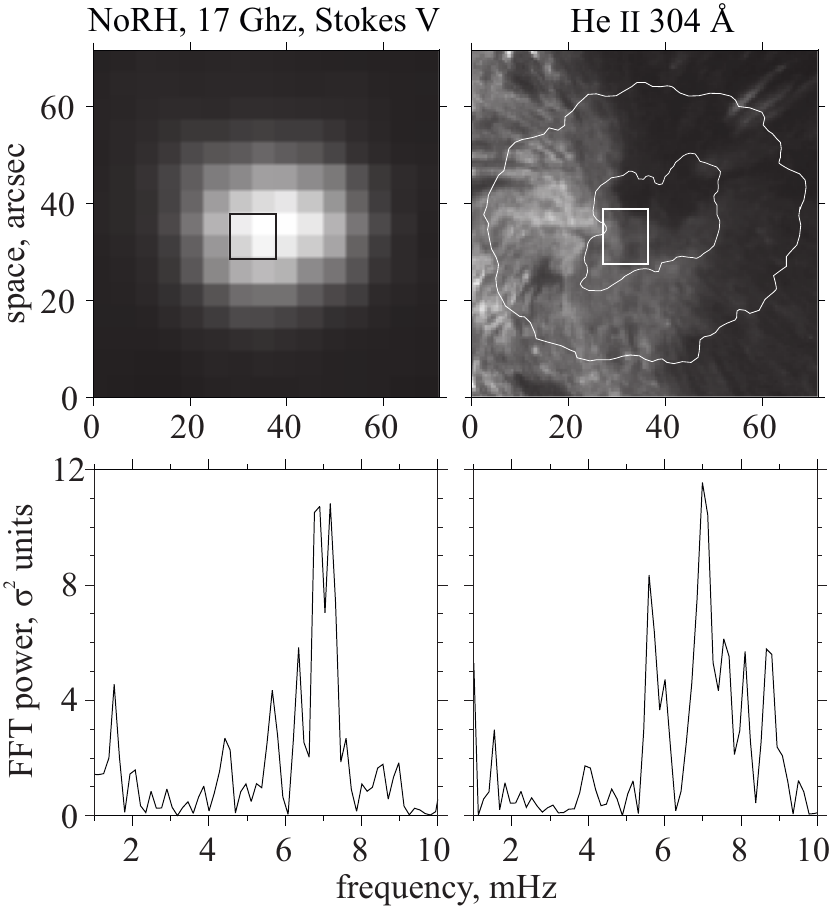}
}
\caption{Left panel: FFT power spectrum of the NoRH 17\,GHz Stokes V signal.
Right panel: FFT power spectrum of the He\,\textsc{ii} 304\,\AA\ intensity.
The signals were averaged over the area marked by the rectangle.
The inner and outer penumbra boundaries are outlined by white solid lines.
        }
\label{fig:norh-304-fft}
\end{figure}

\begin{figure}
\centerline{
\includegraphics[]{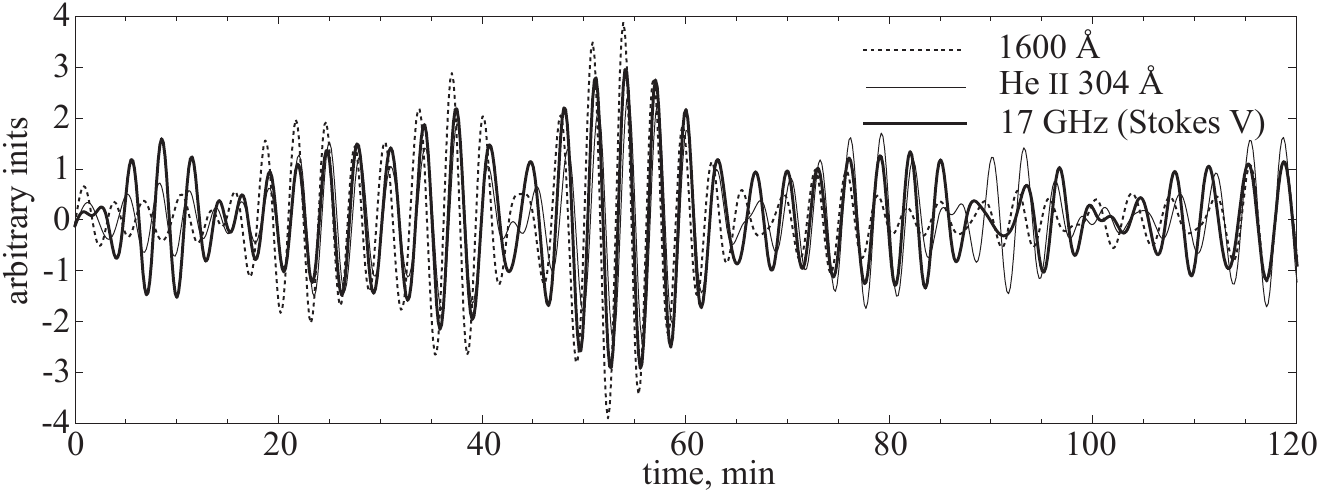}
}
\caption{Intensity variations revealing wave trains in the three-minute band in
the chromospheric and transition region lines 
and the NoRH 17\,GHz Stokes V.
}
\label{fig:wt-1600-304-V}
\end{figure}

\begin{figure}
\centerline{
\includegraphics[]{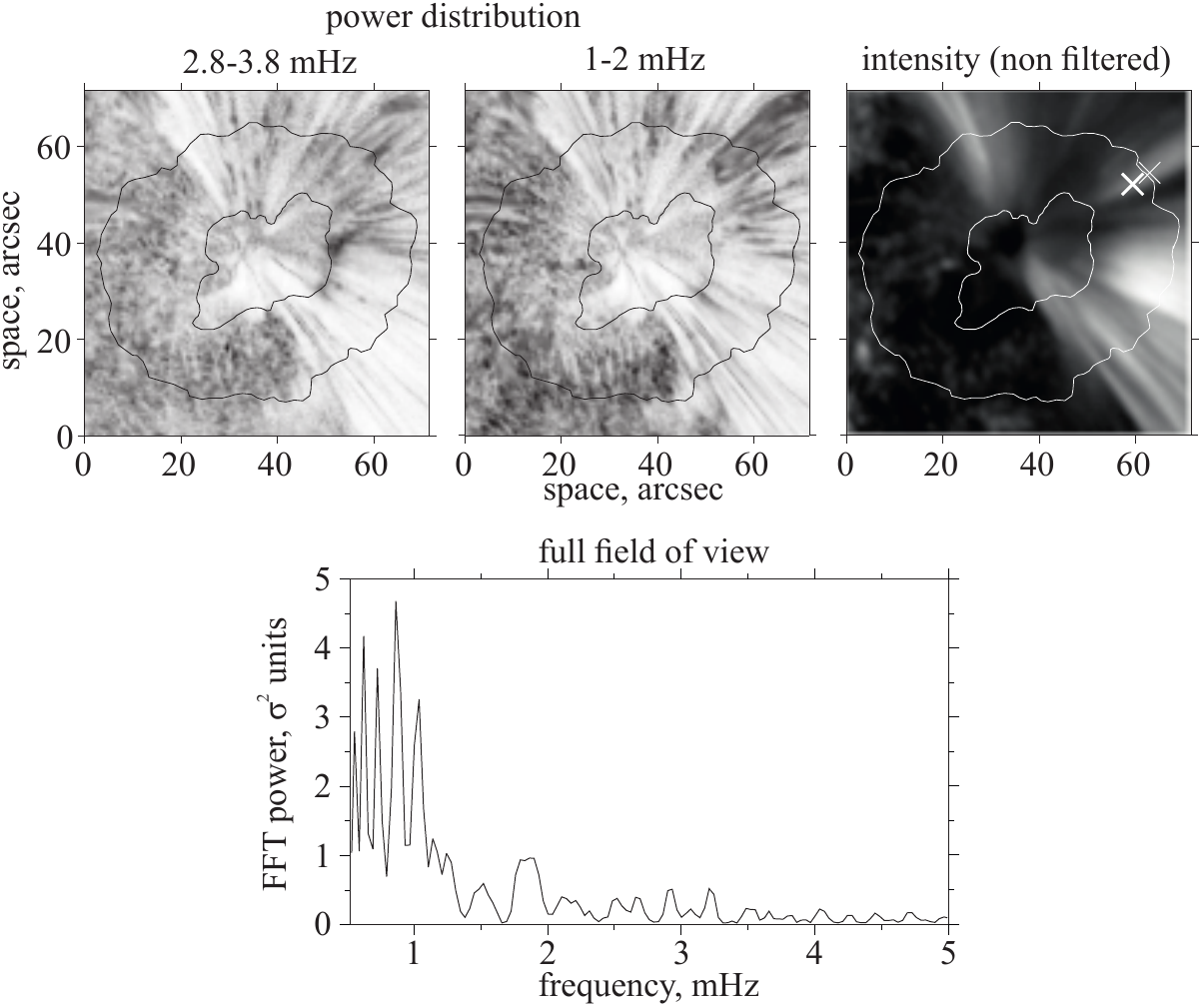}
}
\caption{Fan-like structures in spatial localisations of low-frequency 
coronal oscillations (panels 1 and 2) and inverse sunspot image in the 
Fe\,\textsc{ix} 171\,\AA\ band (panel 3). Panel~4: FFT power spectrum calculated
over the whole field-of-view. The inner and outer penumbra boundaries are 
outlined by black and white solid lines.
}
\label{fig:171-ht-fft}
\end{figure}

\begin{figure}
\centerline{
\includegraphics[]{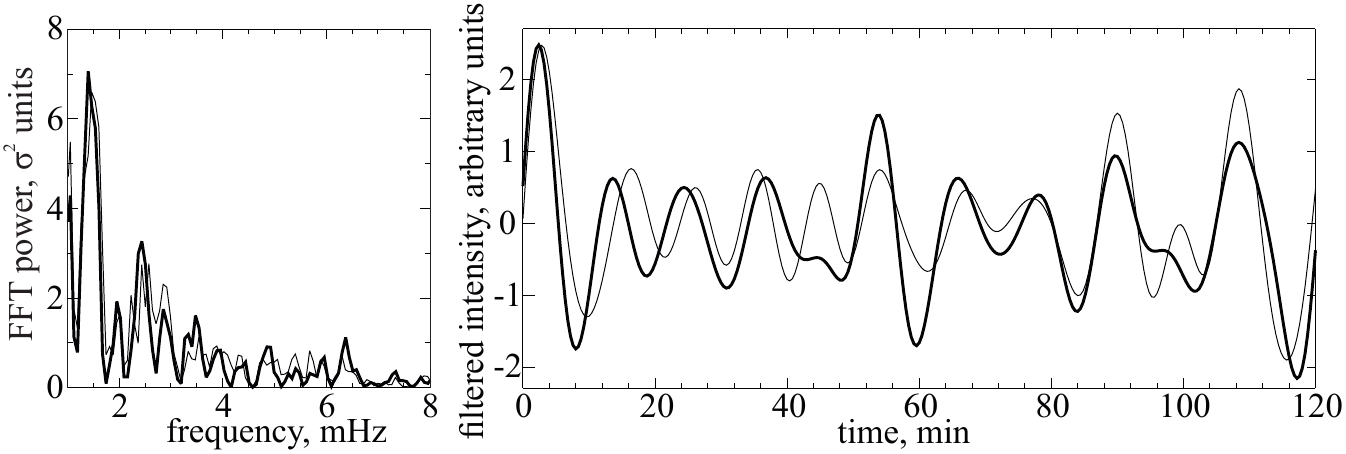}
}
\caption{Oscillation behaviour in coronal loops above sunspot.
Left panel shows the intensity oscillation power spectra in the loop points
marked with crosses in Figure\,\ref{fig:171-ht-fft}. The results
of frequency wavelet filtration are presented in the right panel. The thick
lines show spectra and signals at the points located farther from 
the sunspot centre.
}
\label{fig:171-spectra}
\end{figure}




\par High-frequency waves of the 6\,--\,7\,mHz band concentrate at 
the sunspot central umbral part at all heights and form fragmented 
domains on the power maps (\autoref{fig:ht-6-7}). The power maps were 
produced by using entire time series.
The NoRH at 17\,GHz observation clearly show
three-minute band oscillations in Stokes V polarisation channel
(\autoref{fig:norh-304-fft}).
For the optically thin radiation the line-of-sight magnetic field can
be estimated according to \cite{gel2004ASSL} and is about 730\,G.
Extrapolation of the photospheric magnetic
field to the upper atmosphere gives us the effective height of 2100--2200\,km for the
radio emission that corresponds to the transition region and the field strength of that value.

\par Intensity variations in the narrow band of 3
minute for the 1600 and 304\,\AA\ bands show a wave train structure
that is typical of sunspots. Similar structures are present in
the NoRH 17\,GHz polarisation signal (\autoref{fig:wt-1600-304-V}).
The measured time delay of three-minute propagating waves 
(\autoref{fig:wt-1600-304-V}) from
the level of 1000--1300 km (1600\,\AA) to a height of 17\,GHz emission gives a
value of about 65 seconds, and 5 seconds from the level of 2200\,km (304\,\AA) to a
height of 17\,GHz emission. We can conclude that the calculated 
height of radio emission generation agrees with height estimations for
other spectral lines \citep{avrett2008ApJS,zh1998ApJ} and with the analysis of characteristics
of propagating three-minute waves \citep{rez2012ApJ,kob2013aa}.

\par At the coronal heights, power maps reveal elongated structures 
that outline the coronal loops best at the 1--2 mHz filtration band 
(\autoref{fig:171-ht-fft}). Similarly to the two sunspots analysed in
\cite{kob2015sp}, the 1--2 mHz band oscillations can be
traced along the loops to find the time delay and propagating speed.
For this purpose we find two points along the loop having 
almost identical FFT spectra. These points are marked with
crosses in \autoref{fig:171-ht-fft}. 
Then using the amplitude vs. time plot (\autoref{fig:wt-1600-304-V}), we estimate the 
apparent propagation speed that appears to be $\sim$45\,km/s. The angle between the
line of sight and the considered coronal fan was estimated from the magnetic field
extrapolation and was found to be about 40\degree. Taking into account this value
we find the deprojected wave propagation speed to be about 70\,km/s. This value 
is lower than the expected sound speed in the corona. However, it is consistent with 
other measurements of the slow MHD wave propagation speed in coronal loops, 
ranging from 20\,km/s to 190\,km/s (See Table\,2 in \cite{dem2009ssrv}). 
It is interesting to note that low-frequency oscillations of 1--2 mHz band in sunspots 
can be connected with Evershed flow's pulsations \citep{rimmele1995,kobmak2004}. 
The study of such a possible connection would require 
sunspots located near the solar disc limb. 
Other possible scenarios can be related to the stochastic motion of coronal 
loops or the non-damping transverse oscillations \citep{nis2013aa,anf2013aap}.

\section{Conclusions}
The 5--7 mHz band high frequency oscillations concentrate inside small-scale
elements in umbra and at the umbra-penumbra boundary. These kinds of oscillations
show a well-known wave-train behaviour that also reveals itself in 
the NoRH Stokes V 17\,GHz data. The wave trains observed in UV and radio coincide with 
a cross-correlation coefficient up to 0.75--0.92. The time delays
for the 1600\,\AA, 304\,\AA\ and the NoRH signal agree with the previous results, 
obtained without radio data analysis, and indicate that the radio emission 
effective height is within the 2100--2200\,km range which corresponds to 
the transition region and the magnetic field B$\sim$730\,G.
This value agrees with one measured in He\,{\sc i}\,10830\,\AA\ line 
B$\sim$600--1500\,G, \citep{sol2006asp} and is much higher than the estimated 
magnetic field at the coronal levels B$\sim$10\,G \citep{nak2001aa,zaq2013aa}.
The estimation of the magnetic fields at higher coronal levels would be possible with
availability of new high cadence and frequency radio data from SSRT and CSRH instruments.

\par The 1--2 mHz band low-frequency waves in NOAA\,11711 reproduce the 
behaviour of NOAA\,11479 and NOAA\,11311 analysed by \cite{kob2015sp}:
the power of these waves is concentrated in the annular zone around penumbra 
that expands with height. 
The estimated deprojected propagation speed of the wave train 
travelling along the single fan-loop is about 70\,km/s.

\par The magnetic field inclination angle $[\varphi]$ in the sunspot was estimated in 
two independent ways: from the observations of the dominant oscillations 
frequency and by using the magnetic field extrapolation. 
$[\varphi]$ derived from HMI photospheric data is $\sim$35--40\degree\ for the
umbra-penumbra boundary (12\arcsec\ from the sunspot centre), and corresponds
to the field extrapolation results at the 2200\,km level (\autoref{fig:barycentre}).
$[\varphi]$ is obtained from indirect measurements for He\,{\sc ii}\,304\,\AA\ line is
$\sim$40--45\degree. The inclination estimation results fit in with the general 
structure of a sunspot and provide grounds for new research.

\begin{acks}
We acknowledge the NASA/SDO and NoRH science teams for providing the data.
This work was partly done out on the Solar Data Analysis System operated by the
Astronomy Data Center in cooperation with the Nobeyama Solar Radio Observatory of 
the National Astronomical Observatory of Japan. This study was supported by 
Projects No.\,16.3.2, 16.3.3 of ISTP SB RAS, by the Russian Foundation for
Basic Research under grants No. 12-02-33110 mol\_a\_ved, 15-32-20504 mol\_a\_ved,
15-02-03717-a, 15-02-01077-a, 15-02-01089-a, 13-02-00044-a, 15-02-03835-a. 
We are grateful to anonymous reviewers for their helpful remarks and suggestions.
We acknowledge Y.M.~Kaplunenko for his help in preparing the English version of the paper.
\end{acks}

\bibliographystyle{model2-names}
\bibliography{kolobov}

\end{document}